\documentclass[reprint,amsmath,amssymb,aps,prmaterials,superscriptaddress]{revtex4-2}
\usepackage{graphicx}
\usepackage{siunitx}
\DeclareSIUnit{\hartree}{Ha}

\usepackage{float}
\usepackage[english]{babel}
\usepackage[colorlinks=true, linkcolor=blue, citecolor=blue, urlcolor=blue]{hyperref}
\usepackage{orcidlink}

\makeatletter
\let\origselectlanguage\selectlanguage
\renewcommand{\selectlanguage}[1]{%
  \@ifundefined{l@#1}{\origselectlanguage{english}}{\origselectlanguage{#1}}%
}
\makeatother

\DeclareUnicodeCharacter{03B1}{\ensuremath{\alpha}}
\DeclareUnicodeCharacter{03B2}{\ensuremath{\beta}}
\DeclareUnicodeCharacter{03B3}{\ensuremath{\gamma}}
\DeclareUnicodeCharacter{03B4}{\ensuremath{\delta}}
\DeclareUnicodeCharacter{03BC}{\ensuremath{\mu}}
\DeclareUnicodeCharacter{03C9}{\ensuremath{\omega}}
\DeclareUnicodeCharacter{0394}{\ensuremath{\Delta}}
\DeclareUnicodeCharacter{03A9}{\ensuremath{\Omega}}

\newcommand{\AffLaserHM}{Laser Center HM, Munich University of Applied Sciences HM, Munich, Germany}
\newcommand{\AffNTC}{New Technologies -- Research Centre, University of West Bohemia in Pilsen, Plze\v{n}, Czech Republic}
\newcommand{\AffLiege}{Nanomat, Q-Mat, CESAM, and European Theoretical Spectroscopy Facility, Universit\'e de Li\`ege, Li\`ege, Belgium}
\newcommand{\AffUtrecht}{Institute for Theoretical Physics, Department of Physics, Utrecht University, Utrecht, The Netherlands}
\newcommand{\AffFZU}{FZU -- Institute of Physics of the Czech Academy of Sciences, Prague, Czech Republic}
\newcommand{\AffFNP}{Faculty of Nuclear Sciences and Physical Engineering, Czech Technical University in Prague, Prague, Czech Republic}
\newcommand{\TiAlloy}{\mbox{Ti-6Al-4V}}
\begin{document}

%% --- TITLE ---
\title{First-principles electronic transport properties of Ti and \TiAlloy{}
for modeling ultrashort-pulse laser ablation}

%% --- AUTHORS & AFFILIATIONS ---
\author{Korbinian Hobmaier\,\orcidlink{0009-0009-6110-224X}}
\affiliation{\AffLaserHM}

\author{Guillaume E. Allemand\,\orcidlink{0000-0002-9080-9799}}
\affiliation{\AffLiege}

\author{Alberto Marmodoro\,\orcidlink{0000-0003-4174-9643}}
\affiliation{\AffNTC}
\affiliation{\AffFZU}
\affiliation{\AffFNP}

\author{Matthieu J. Verstraete\,\orcidlink{0000-0001-6921-5163}}
\affiliation{\AffLiege}
\affiliation{\AffUtrecht}

\author{J\'an Min\'ar\,\orcidlink{0000-0001-9735-8479}}
\email{jminar@ntc.zcu.cz}
\affiliation{\AffNTC}

\author{Heinz P. Huber\,\orcidlink{0000-0003-2444-9833}}
\email{heinz.huber@hm.edu}
\affiliation{\AffLaserHM}
\affiliation{\AffNTC}

\author{David Redka\,\orcidlink{0000-0002-7306-2232}}
\email{dredka@hm.edu}
\affiliation{\AffLaserHM}

\date{\today}

%% --- ABSTRACT ---
\begin{abstract}
Predictive modeling of ultrashort-pulse laser ablation requires temperature-dependent material parameters derived from the electronic structure, namely the electronic thermal conductivity,
electron--phonon coupling, and heat capacity. These parameters are well documented for elemental metals but remain sparsely documented for alloys, apart from application-relevant exceptions such as stainless steels.
The technologically important titanium alloy \TiAlloy{} is a prominent example, which is still modeled using elemental-titanium values. We compute the electronic transport of hcp Ti and \TiAlloy{} from first principles, using the Kubo--Greenwood formalism within the Korringa--Kohn--Rostoker coherent-potential-approximation framework,
treating chemical and thermal disorder on equal footing. For elemental Ti, the calculated electrical resistivity agrees with independent \textsc{abinit} electron--phonon calculations and experiment, and also
reproduces the high-temperature saturation near the Mott--Ioffe--Regel limit. Under electron--phonon nonequilibrium, the electronic thermal conductivity saturates and then decreases
with electronic temperature, reaching a maximum of about \SI{2.97}{\kilo\watt\per\metre\per\kelvin} in Ti but only \SI{0.47}{\kilo\watt\per\metre\per\kelvin} in \TiAlloy{}, a factor of 6.4 lower. In two-temperature-model simulations the alloy and elemental
parameter sets yield peak lattice temperatures differing by only about 1.4\%, because reduced electron--phonon coupling and thermal conductivity offset each other,
consistent with reported experimental ablation thresholds that differ by about 3\%, well within their measurement uncertainties. This compensation is strongest for sub-picosecond pulses and weakens across the \SI{1}{ps} to \SI{10}{ps} crossover toward longer-pulse behavior. 
Replacing the first-principles thermal conductivity with the low-temperature Drude limit shifts the peak lattice temperature by up to 19\%, showing that the functional form of the transport model is even more important than
the elemental vs alloy distinction for predictive accuracy.
\end{abstract}

\maketitle

%% --- BODY ---
%% (section inputs to be added)
\section{Introduction}

\TiAlloy{} (grade~5 titanium) is the workhorse of titanium applications in medicine, science and industry, valued for its high strength-to-weight ratio, corrosion resistance, biocompatibility, and thermal stability~\cite{marin_biomedical_2023,abd-elaziem_titanium-based_2024}, which makes it central to, e.g. lightweight aerospace design and to orthopedic and dental implants~\cite{liu_review_2021,yin_crystal_2022,srivastava_additive_2024,ezugwu_overview_2003}. These applications place strict requirements on machining precision and surface texture, for which laser processing has become a key enabling technology. Ultrashort-pulse (USP) lasers enable micrometer-scale machining and sub micrometer functional surface textures that improve wettability, osseointegration, and cellular response~\cite{ozan_laser_2025,cunha_human_2015,liu_characterization_2021,gnilitskyi_enhanced_2024,garcia-hernandez_effect_2025}.

For pulse durations shorter than the electron--phonon equilibration time (in the ps range), predictive modeling relies on the two-temperature model (TTM)~\cite{anisimov_electron_1974}, which couples electronic and lattice subsystems through Fourier heat equations, linked by electron--phonon energy transfer~\cite{rethfeld_modelling_2017,allen_theory_1987}. The TTM for metals and alloys is based on four electron ($T_\mathrm{e}$) and lattice ($T_\mathrm{l}$) temperature-dependent material parameters, namely the electronic and lattice heat capacities $C_{\mathrm{e}}$ and $C_{\mathrm{l}}$, the electronic thermal conductivity $\kappa_{\mathrm{e}}$, and the electron--phonon coupling factor $G$. The lattice thermal conductivity, which scales as $1/T_{\mathrm{l}}$ and remains smaller than the electronic conductivity, is typically disregarded for metals and alloys.

For pure metals these parameters are well documented. Lin \textit{et~al.} provided an electron-temperature-dependent description of the electron--phonon coupling $G$ from the electronic density of states~\cite{lin_electron-phonon_2008}. $\kappa_{\mathrm{e}}$ is usually captured by the low-temperature Drude limit, $\kappa_{\mathrm{e}} = \kappa_{\mathrm{e0}}\,T_{\mathrm{e}}/T_{\mathrm{l}}$ \cite{rethfeld_modelling_2017}, where $\kappa_{\mathrm{e0}}$ is the electronic thermal conductivity at room temperature, or by more complex models~\cite{petrov_thermal_2015}. For alloys, however, the data are far more sparse. Winter \textit{et~al.}~\cite{winter_temperature_2016} and B\'evillon \textit{et~al.}~\cite{bevillon_ab_2015} extended the density-of-states-based $C_{\mathrm{e}}$ and $G$ description to stainless steel, and first-principles studies have addressed the equilibrium electronic transport of concentrated solid solutions~\cite{samolyuk_temperature_2018,samolyuk_electronphonon_2016}. The electronic thermal conductivity under electron--phonon nonequilibrium, by contrast, remains largely undetermined apart from a few exceptions~\cite{medvedev_stainless_2025}, and especially so for \TiAlloy{}. Most USP ablation simulations of \TiAlloy{} therefore default to elemental-titanium parameters, in particular for $\kappa_{\mathrm{e}}$ and $G$~\cite{kiran_kumar_theoretical_2019,peng_numerical_2023,chen_multiphysics_2024,li_electromagnetic_2025}.

In the present work we address these gaps. The electronic structure and transport properties of the hexagonal phases of Ti and \TiAlloy{} are computed as functions of both $T_{\mathrm{e}}$ and $T_{\mathrm{l}}$ within the Korringa--Kohn--Rostoker (KKR) Green's function formalism~\cite{ebert_calculating_2011}, treating chemical disorder by the coherent potential approximation (CPA) and thermal disorder by the alloy-analogy model on the same footing. For elemental Ti, the resistivity obtained from the Kubo--Greenwood linear-response framework~\cite{greenwood_boltzmann_1958} is cross-checked against independent \textsc{abinit} electron--phonon calculations~\cite{verstraete_abinit_2025}. The heat capacity $C_{\mathrm{e}}$, the coupling factor $G$, and, in particular, $\kappa_{\mathrm{e}}$ under electron--phonon nonequilibrium are then computed for both Ti and the alloy. These properties feed one-dimensional TTM simulations, in which the full first-principles parameters are compared with the approximations commonly used for elemental Ti and with the corresponding \TiAlloy{} values. Finally, a sensitivity analysis quantifies how each parameter influences the prediction of the ablation threshold as a function of pulse duration.
\section{Methods}
\label{sec:methods}

\subsection{Electronic structure calculations}
\label{sec:DFT}

Electronic-structure calculations for Ti and \TiAlloy{} were performed within the fully relativistic KKR Green-function formalism as implemented in SPR-KKR~\cite{korringa_calculation_1947,ebert_calculating_2011}. 
Exchange and correlation were described within the generalized gradient approximation in the Perdew--Burke--Ernzerhof (PBE) parametrization~\cite{perdew_generalized_1996}. 
The angular-momentum expansion of the Green function was truncated after $f$-character contributions, 
and Brillouin-zone (BZ) integrations employed a $60\times60\times33$ $k$-point mesh for the hexagonal lattice.

For \TiAlloy{}, substitutional chemical disorder among Ti, Al, and V was treated within the CPA, in which each site carries the average occupation with atomic fractions $x_{\mathrm{Ti}}=0.862$, $x_{\mathrm{Al}}=0.102$, and $x_{\mathrm{V}}=0.036$,
corresponding to the nominal composition of 6\,wt.\,\%\,Al and 4\,wt.\,\%\,V. The \TiAlloy{} results should therefore be understood as an idealized, substitutionally 
disordered reference alloy. The experimental lattice parameters (in-plane
$a$ and the out-of-plane $c$) of the hexagonal close-packed (hcp) structure are listed in Table~\ref{tab:params}.

\begin{table}[t]
  \caption{
      Structural and thermal input parameters. The lattice constants are room-temperature values,
      averaged over the references cited for pure Ti~\cite{lutjering_titanium_2007,lonardelli_situ_2007,wood_lattice_1962}
      and for \TiAlloy{}~\cite{montanari_lattice_2008,swarnakar_thermal_2011, lonardelli_situ_2007, xu_martensitic_2018}.
      The mechanical Debye temperature $\Theta_{\mathrm{D}}^{\mathrm{mech}}$, taken from Chen and Sundman~\cite{chen_calculation_2001} and applied to \TiAlloy{} as well (consistent with the nearly identical 
      Sandia hcp-phase values~\cite{kerley_equations_2003}), enters alloy-analogy-model (AAM) thermal-displacementcalculation,
      whereas the effective thermodynamic Debye temperature $\Theta_{\mathrm{D}}^{\mathrm{th}}$ enters the electron--phonon coupling $G$ and the lattice heat capacity $C_{\mathrm{l}}$ (see the Supplemental Material).}
  \label{tab:params}
  \begin{ruledtabular}
    \begin{tabular}{lcccc}
      System & $a$ (\AA) & $c$ (\AA) & $\Theta_{\mathrm{D}}^{\mathrm{mech}}$ (K) & $\Theta_{\mathrm{D}}^{\mathrm{th}}$ (K) \\
      \hline
      Ti & 2.95 & 4.68 & 385 & 415 \\
      \TiAlloy{} & 2.93 & 4.68 & 385 & 415 \\
    \end{tabular}
  \end{ruledtabular}
\end{table}

Lattice thermal disorder was incorporated through the alloy-analogy model (AAM)~\cite{ebert_calculating_2015},
in which thermally displaced atomic configurations are mapped onto an effective substitutional alloy. The displaced-alloy components were generated from a symmetry-adapted cluster
construction that respects the crystal symmetry, with 18 isotropic directions,
and the displacement distribution was taken to be isotropic.
Test calculations using molecular-dynamics and first-principles data showed only negligible anisotropy
(below 5\,\% relative difference in the mean-square atomic displacement) over the temperature range relevant here
(see Supplemental Material).

The temperature-dependent mean-square displacement was obtained from the Debye model~\cite{solyom_fundamentals_2009},
\begin{equation}
  \langle u^2 \rangle = \frac{9\hbar^2}{k_{\mathrm{B}} \Theta_{\mathrm{D}}^{\mathrm{mech}} M} \left[\frac{1}{4} + \left(\frac{T}{\Theta_{\mathrm{D}}^{\mathrm{mech}}}\right)^2 \int_{0}^{\Theta_{\mathrm{D}}^{\mathrm{mech}}/T} \frac{t}{e^t-1}\,\mathrm{d}t \right],
  \label{eq:msd}
\end{equation}
where $M$ is the mean atomic mass and $\Theta_{\mathrm{D}}^{\mathrm{mech}}$
the mechanical Debye temperature (listed in Table~\ref{tab:params}).
The zero-point contribution was omitted following Ref.~\cite{ebert_calculating_2015}.
First-principles calculations performed with the \textsc{abinit} package (see Sec.~\ref{sec:transport}) yield a consistent value of $360\text{ K}$ (see Supplemental Material).

Initial tests showed that thermal expansion and self-consistent displaced potentials change the 
temperature-dependent resistivity only marginally (see the Supplemental Material).
All subsequent calculations were therefore carried out with the experimental lattice parameters of Table~\ref{tab:params} and with the self-consistent potential of the corresponding undistorted structure.
Finite electronic-temperature was considered in the subsequent transport and thermodynamic quantities through the Fermi occupations and the temperature-dependent chemical potential, while the underlying self-consistent potential
was not recalculated at finite electronic temperature. Apart from a chemical-potential shift the density-of-states (DOS) shape changes only marginally over the relevant range (see the Supplemental Material).

\subsection{Electronic transport properties}
\label{sec:transport}

Electronic transport was evaluated within the Kubo--Greenwood (KG) linear-response formalism as implemented in SPR-KKR~\cite{ebert_calculating_2011,greenwood_boltzmann_1958}. The energy-resolved conductivity tensor is
\begin{equation}
  \sigma_{\mu\nu}(E)
  =
  \frac{\hbar}{\pi N \Omega}\,
  \mathrm{Tr}\,
  \bigl\langle
  \hat{j}_{\mu}\,\mathrm{Im}\,G^{+}(E)\,
  \hat{j}_{\nu}\,\mathrm{Im}\,G^{+}(E)
  \bigr\rangle_{c}\,,
  \label{eq:KG}
\end{equation}
where $N$ is the number of atomic sites, $\Omega$ the unit-cell volume, $\hat{j}_{\mu}$ the relativistic current-density operator,
$G^{+}(E)$ the retarded single-particle Green function, and $\langle\cdots\rangle_{c}$ the CPA configurational average, including 
vertex corrections~\cite{ebert_calculating_2011}. Equation (\ref{eq:KG}) retains only the Fermi-surface term of the full Kubo--Bastin conductivity~\cite{kodderitzsch_linear_2015}, and the electronic-temperature dependence enters solely through the Fermi--Dirac occupation in Eq.~(\ref{eq:GTC}).
The full BZ was sampled with a $56\times56\times30$ \textbf{k}-point mesh after convergence tests (see the Supplemental Material). 
All transport results reported below were obtained by including angular momentum components up to $f$ orbitals.
The dependence on the angular-momentum cutoff is documented in the Supplemental Material.

In this scheme, the lattice temperature $T_{\mathrm{l}}$ enters through the AAM and the electronic temperature $T_{\mathrm{e}}$ 
through the Fermi--Dirac occupations in the transport integrals.
The energy-resolved conductivity $\sigma_{\mu\nu}(E)$ was computed for a discrete set of lattice temperatures up to $T_{\mathrm{l}}=\SI{2100}{\kelvin}$
and then linearly extrapolated to $T_{\mathrm{l}}=\SI{2500}{\kelvin}$ ($\approx 1.3\,T_{\mathrm{m}}$). At each $T_{\mathrm{l}}$,
the energy grid covered the interval $\pm\,6\,k_{\mathrm{B}} T_{\mathrm{e,max}}$ around the Fermi level with 50 sampling points,
where $T_{\mathrm{e,max}}=\SI{50}{\kilo\kelvin}$. The electronic thermal conductivity was then obtained from the generalized (Onsager) transport coefficients (GTC)~\cite{jonson_motts_1980,samolyuk_temperature_2018, ricci_ab_2017},
\begin{equation}
  L^{\mu\mu}_{ij}
  =
  (-1)^{i+j}
  \int_{-\infty}^{\infty}
  \mathrm{d}E\;
  \sigma_{\mu\mu}(E)\,
  (E-\mu_{\mathrm{c}})^{i+j-2}\,
  \biggl(-\frac{\partial f}{\partial E}\biggr),
  \label{eq:GTC}
\end{equation}
where $f \equiv f(E,\mu_{\mathrm{c}},T_{\mathrm{e}})$ is the Fermi--Dirac distribution at electronic temperature $T_{\mathrm{e}}$, and $\mu_{\mathrm{c}}(T_{\mathrm{e}})$ is the $T_{\mathrm{e}}$ dependent chemical potential calculated from charge conservation of total valence-electron count from the DOS \cite{winter_temperature_2016}. The longitudinal electronic thermal conductivity follows as
\begin{equation}
  \kappa_{\mu\mu}(T_{\mathrm{e}},T_{\mathrm{l}})
  =
  \frac{1}{e^{2}\,T_{\mathrm{e}}}
  \biggl(
    L^{\mu\mu}_{22}
    -
    \frac{L^{\mu\mu}_{12}\,L^{\mu\mu}_{21}}%
    {L^{\mu\mu}_{11}}
  \biggr).
  \label{eq:kappa_e}
\end{equation}

As an independent benchmark, the electrical resistivity of hcp Ti was computed with the electron--phonon coupling module of \textsc{abinit} version 10.5.8~\cite{verstraete_abinit_2025} by solving the iterative Boltzmann transport equation~\cite{giustino_electron-phonon_2017}. Fully relativistic norm-conserving PBE pseudopotentials from the PseudoDojo library~\cite{van_setten_pseudodojo_2018}, including spin--orbit coupling, were employed. The plane-wave cutoff was set to \SI{40}{\hartree}, with a $\Gamma$-centered $10\times10\times10$ \textbf{k}-point grid and a $5\times5\times5$ \textbf{q}-point grid for the phonon perturbations. Whereas the SPR-KKR calculations used the experimental lattice parameters of Table~\ref{tab:params}, the ABINIT reference cell was fully relaxed, yielding $a=\SI{2.943}{\angstrom}$ and $c=\SI{4.640}{\angstrom}$ (residual stresses below \SI{6e-3}{\giga\pascal}). An energy window of \SI{0.4}{\electronvolt} around the Fermi level and an interpolated $40\times40\times40$ \textbf{q}-point grid were used for the scattering potentials. Full methodological details are given in Refs. \cite{allemand_first-principles_2025,brunin_electron-phonon_2020,brunin_phonon-limited_2020}.

\subsection{Two-temperature model}
\label{sec:TTM}

The thermal response to USP laser irradiation was modeled with a one-dimensional TTM~\cite{anisimov_electron_1974},
initialized at $T_{\mathrm{e}}=T_{\mathrm{l}}=\SI{300}{\kelvin}$. Under stress confinement, USP laser ablation proceeds through 
photomechanical spallation and, at higher fluence, phase explosion~\cite{paltauf_photomechanical_2003,wu_microscopic_2014}.
Mechanical and hydrodynamic response as well as phase change are not considered.
The peak lattice temperature reached during the simulation serves as a proxy for the ablation threshold, as it sets the deposited energy and the associated buildup of thermoelastic stress that drives material removal. The coupled energy-balance equations for the electronic and lattice subsystems read

\begin{equation}
  C_{\mathrm{e}}\,\frac{\partial T_{\mathrm{e}}}{\partial t}
  =
  \frac{\partial}{\partial z}
  \biggl[\kappa_{\mathrm{e}}(T_{\mathrm{e}},T_{\mathrm{l}})\,
  \frac{\partial T_{\mathrm{e}}}{\partial z}\biggr]
  -
  G(T_{\mathrm{e}})\,(T_{\mathrm{e}}-T_{\mathrm{l}})
  +
  S(z,t)\,,
  \label{eq:TTM_e}
\end{equation}
\begin{equation}
  C_{\mathrm{l}}\,\frac{\partial T_{\mathrm{l}}}{\partial t}
  =
  \frac{\partial}{\partial z}
  \biggl[\kappa_{\mathrm{l}}(T_{\mathrm{l}})\,
  \frac{\partial T_{\mathrm{l}}}{\partial z}\biggr]
  +
  G(T_{\mathrm{e}})\,(T_{\mathrm{e}}-T_{\mathrm{l}})\,,
  \label{eq:TTM_l}
\end{equation}
where $z$ is the depth coordinate normal to a laser irradiated surface.

The electronic heat capacity was computed from the self-consistent DOS, $n(E)$, according to
\begin{equation}
  C_{\mathrm{e}}(T_{\mathrm{e}})
  =
  \int_{-\infty}^{\infty}\mathrm{d}E\;
  (E-\mu_{\mathrm{c}})\,
  \frac{\partial f(E,\mu_{\mathrm{c}},T_{\mathrm{e}})}{\partial T_{\mathrm{e}}}\;
  n(E)\,,
  \label{eq:Ce}
\end{equation}
with the chemical potential $\mu_{\mathrm{c}}(T_{\mathrm{e}})$ fixed by the same valence-electron conservation as in Eq.~(\ref{eq:GTC}). The electron--phonon coupling factor was evaluated following the formalism of Lin \textit{et~al.}~\cite{lin_electron-phonon_2008},
\begin{equation}
  G(T_{\mathrm{e}})
  =
  \frac{\pi\hbar k_{\mathrm{B}}\,\lambda\langle\omega^{2}\rangle}%
  {n(E_{\mathrm{F}})}
  \int_{-\infty}^{\infty}\mathrm{d}E\;
  n^{2}(E)\,
  \biggl(-\frac{\partial f}{\partial E}\biggr),
  \label{eq:Gep}
\end{equation}
where $n(E_{\mathrm{F}})$ is the DOS at the Fermi level. The product $\lambda\langle\omega^{2}\rangle$ is defined through the Eliashberg spectral function~\cite{allen_neutron_1972} as $\lambda\langle\omega^{2}\rangle = 2\!\int_{0}^{\infty}\!\mathrm{d}\omega\,\alpha^{2}\!F(\omega)\,\omega$. Approximating $\alpha^{2}\!F(\omega)$ by a Debye spectrum with frequency-independent electron--phonon matrix elements yields $\langle\omega^{2}\rangle = \tfrac{1}{2}\,\bigl(k_{\mathrm{B}}\Theta_{\mathrm{D}}^{\mathrm{th}}/\hbar\bigr)^{2}$, evaluated with the thermodynamic Debye temperature $\Theta_{\mathrm{D}}^{\mathrm{th}}$ of Table~\ref{tab:params}, following Lin \textit{et~al.}~\cite{lin_electron-phonon_2008}. The electron--phonon coupling constants $\lambda$ are taken from Ref.~\cite{mcmillan_transition_1968}. For \TiAlloy{}, effective values were constructed from composition-weighted elemental data following Ref.~\cite{winter_temperature_2016}, giving $\lambda_\mathrm{Ti} = 0.38$ and $\lambda_{\text{Ti-6Al-4V}} = 0.388$. As an independent check, the \textsc{abinit} electron--phonon calculation (Sec.~\ref{sec:transport}) yields $\lambda = 0.548$ for elemental Ti directly from the Eliashberg spectral function, larger than the tabulated McMillan value. The product $\lambda\langle\omega^{2}\rangle$, the quantity that actually enters Eq.~(\ref{eq:Gep}), nevertheless agrees to within about 11\,\% (\SI{216.4}{\milli\electronvolt\squared} from \textsc{abinit} vs.\ \SI{243}{ \milli\electronvolt\squared} from the Debye-spectrum estimate), supporting the present parametrization. As the experimental Debye temperatures of Ti and \TiAlloy{} are nearly identical (Table~\ref{tab:params}) and the mean atomic mass changes by only about 4\,\%, $\lambda\langle\omega^2\rangle$ is essentially material-independent, so the computed difference in $G$ is of electronic origin.

The lattice heat capacity $C_{\mathrm{l}}(T_{\mathrm{l}})$ and 
lattice thermal conductivity $\kappa_{\mathrm{l}}(T_{\mathrm{l}})$ were parametrized from experimental data.
Functional forms and references are given in the Supplemental Material.

The laser source term was modeled as a Gaussian temporal pulse with Beer--Lambert linear absorption in z-dimension,
\begin{equation}
  \begin{split}
  S(z,t)
  ={} &
  \frac{(1-R)\,F}{d_{\mathrm{opt}}}\,
  \sqrt{\frac{4\ln 2}{\pi\,\tau_{\mathrm{p}}^{2}}}\\
  & \times
  \exp\!\biggl(\!-\frac{z}{d_{\mathrm{opt}}}\biggr)\,
  \exp\!\biggl(\!-4\ln 2\,\frac{(t-t_{0})^{2}}{\tau_{\mathrm{p}}^{2}}\biggr),
  \end{split}
  \label{eq:source}
\end{equation}
where $F$ is the incident peak fluence, $R$ the normal-incidence reflectance, 
$d_{\mathrm{opt}}$ the optical penetration depth at $\lambda_{\mathrm{L}}=\SI{1030}{\nano\meter}$,
and $\tau_{\mathrm{p}}=\SI{300}{\femto\second}$ the temporal full width at half maximum of the pulse intensity. 
The optical constants were taken identical for Ti and \TiAlloy{},
with $R=0.614$ and $d_{\mathrm{opt}}=\SI{20.4}{\nano\meter}$, computed from the extinction coefficient 
$k = 3.99$ via $d_\mathrm{opt} = \lambda / (4\pi k)$~\cite{johnson_optical_1974}. 
Transient changes of the optical constants during the pulse were neglected, as the reflectivity of metals remains approximately
constant up to the ablation threshold fluence~\cite{winter_ultrashort_2021}.

The simulation domain extended \SI{1}{\micro\meter} into the bulk with grid spacing of $\SI{1}{\nano\meter}$. 
Zero-flux boundary conditions were imposed at the surface and at the rear boundary.
The coupled equations were integrated numerically with \textsc{comsol} Multiphysics version~6.4~\cite{noauthor_comsol_2024}.

The KKR formalism and the AAM are formally defined for crystalline solids. Lattice temperatures above about $1.3\,T_{\mathrm{m}}$,
the estimated threshold for ultrafast homogeneous melting under isochoric conditions~\cite{rethfeld_ultrafast_2002},
therefore extend the present treatment beyond its strict domain of validity. These data were retained as a first-order approximation,
since the electronic DOS of close-packed metals changes only weakly upon melting~\cite{jank_electronic_1991}.

\section{Results and discussion}
\subsection{Equilibrium electronic transport in Ti}
\label{sec:transport_Ti}

\begin{figure*}[ht!]
    \centering
    \includegraphics[width=\textwidth]{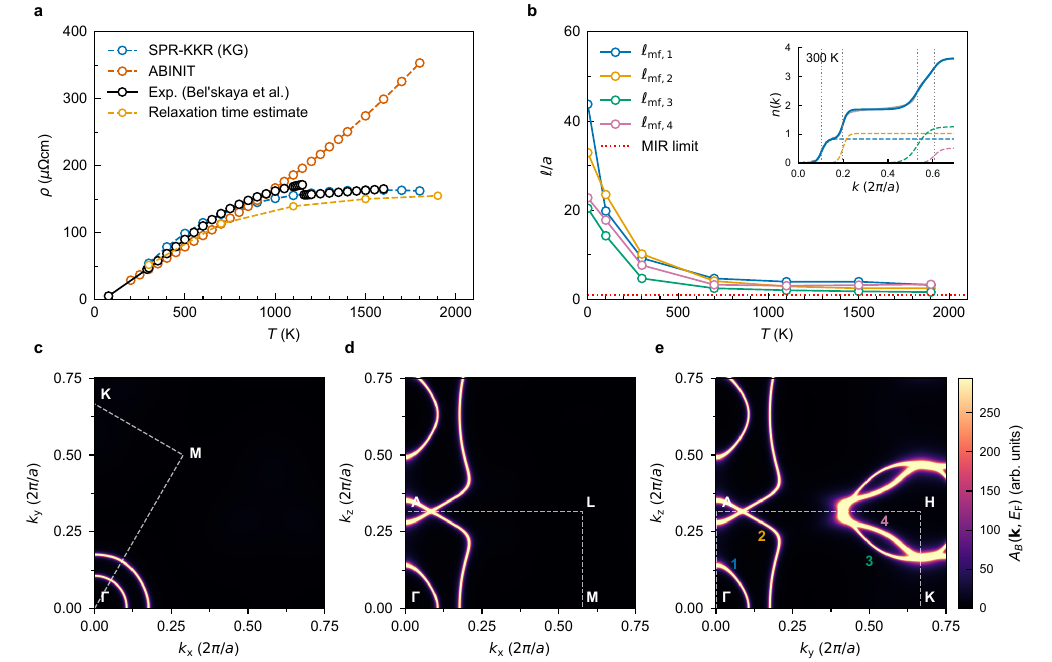}
    \caption{
    Electronic transport and Fermi surface of hcp Ti from first-principles calculations.
    (a) Temperature-dependent electrical resistivity, showing the isotropic average from KKR-CPA KG calculations within SPR-KKR (blue), \textsc{abinit} results (orange),
    the relaxation time estimate from Eq.~(\ref{eq:boltzmann}) (yellow),
    and experimental data from Bel'skaya \textit{et~al.}~\cite{belskaya_experimental_2005} (black).
    (b) Orientationally averaged, sheet-resolved mean free path $\ell_\mathrm{mf, i}$ (in units of the lattice constant $a$) as a function of temperature, 
    extracted from hyperbolic-tangent fits to the momentum distribution $n(k)$ along seven special directions~\cite{kontrym-sznajd_how_2015}. 
    Four Fermi-surface sheets are resolved: $\ell_\mathrm{mf, 1}$ (blue), $\ell_\mathrm{mf, 2}$ (yellow), $\ell_\mathrm{mf, 3}$ (green), and $\ell_\mathrm{mf, 4}$ (pink). 
    The horizontal dotted line marks the Mott--Ioffe--Regel (MIR) limit, $\ell_\mathrm{mf} = a$. The inset shows the computed $n(k)$ at 300\,K for one representative path together with 
    the fit (blue) and the individual sheet contributions (dashed), with vertical dotted lines indicating the Fermi wave vectors.
    (c--e) Bloch spectral function $A_B(\mathbf{k}, E_F)$ at the Fermi energy, shown as plane cuts through the $\Gamma$ point without thermal disorder (no AAM).}
    \label{Fig:aTi_eq_transport}
\end{figure*}

Figure~\ref{Fig:aTi_eq_transport}(a) compares the calculated resistivity of hcp Ti with the experimental data of 
Bel'skaya \textit{et~al.}~\cite{belskaya_experimental_2005}.
Both computational approaches, the KG calculation within SPR-KKR 
($T_{\mathrm{e}}=0$\,K, since electron--electron scattering $\ll$ electron--phonon 
scattering in equilibrium~\cite{rethfeld_modelling_2017}) and the \textsc{abinit} 
electron--phonon calculation, reproduce the measured resistivity well at moderate 
temperatures, with values at 300\,K of 54.3 $\mu\Omega$\,cm and 45.3\,$\mu\Omega$\,cm, respectively, 
compared with the experimental 47.5\,$\mu\Omega$\,cm, and with mutual agreement 
between the two methods up to about 700\,K.
Above this range the two methods diverge. The \textsc{abinit} result continues 
to rise over the entire temperature range, as expected from the standard diffusive transport picture,
%perturbative treatment of electron--phonon scattering in density-functional perturbation theory
which yields a linearly increasing 
scattering rate at high $T$. The experimental resistivity, in contrast, deviates from linearity above about 700\,K 
and saturates close to the hcp--bcc (body-centered cubic) phase transition at about 1155\,K, across which the measured resistivity drops by roughly 5\,\%. 
The KG calculation reproduces this saturation. The bcc phase above 1155\,K was not included in either calculation, so the computed saturation 
arises entirely within the hcp CPA--AAM framework and is not associated with the structural transition.

The saturation of the KG resistivity is attributed to the progressive loss of quasiparticle coherence with increasing thermal disorder 
within the CPA--AAM framework. To quantify this, the quasiparticle coherence length $\xi_i$ of each Fermi-surface sheet 
was extracted from the \textbf{k}-resolved occupation number $n(\mathbf{k})$, following the procedure of Robarts \textit{et~al.}~\cite{robarts_extreme_2020}.
The momentum distribution was obtained by integrating the Bloch spectral function (BSF, $A_B(\mathbf{k},E)$) up to the Fermi energy,
\begin{equation}
  n(\mathbf{k}) = \int_{-\infty}^{E_F} A_B(\mathbf{k},E)\,\mathrm{d}E\,.
  \label{eq:nk}
\end{equation}
For an ordered crystal, $n(\mathbf{k})$ exhibits sharp unit-step discontinuities at each Fermi-surface crossing, which are progressively smeared by thermal disorder. Each crossing was fitted with a hyperbolic tangent step function~\cite{robarts_extreme_2020},
\begin{equation}
  n(\mathbf{k}) = \sum_{i=1}^{N}\frac{c_i}{2}
  \left[1+\tanh\!\left(\frac{4\,\xi_i\,(\mathbf{k}-\mathbf{k}_{\mathrm{F},i})}{a}\right)\right] + C\,,
  \label{eq:tanh}
\end{equation}
where $k_{\mathrm{F},i}$ is the Fermi wave vector of the $i$-th sheet, $\xi_i$ the coherence length of that sheet along the given direction,
$c_i$ the step amplitude, and $C$ a constant offset.
The Fermi wave vectors were determined from a free fit at 0\,K and held fixed at finite temperature,
consistent with Luttinger's theorem~\cite{luttinger_fermi_1960}.
The momentum distribution was sampled along seven special directions in the hcp BZ according to Ref.~\cite{kontrym-sznajd_how_2015}.
Fermi surface averaging of $\xi_i$ over these directions gives the sheet-resolved electronic mean free path $\ell_{\mathrm{mf}, i}$.
Individual paths are documented in the Supplemental Material.

Figure~\ref{Fig:aTi_eq_transport}(b) shows the sheet-resolved mean free path $\ell_{\mathrm{mf}, i}/a$ for the four resolved Fermi surface sheets as a function of temperature.
The mean free paths of all sheets decrease rapidly with temperature and saturate above about 1100\,K at $\ell_{\mathrm{mf}, i}/a \approx 3$,
corresponding to a mean free path of a few lattice constants, within a small factor of the Mott--Ioffe--Regel (MIR) limit $\ell_\mathrm{mf} \approx a$.
The finite values at $\SI{0}{\kelvin}$, where no thermal disorder enters the AAM and the lattice is fixed,
are an artifact of the finite imaginary energy $\mathrm{Im}\,E = 1$\,mRy used in the spectral-function evaluation.
The saturation of $\ell_\mathrm{mf}$ within the KKR-CPA-AAM formalism directly explains the resistivity saturation observed in panel~(a).
The inset shows the four tanh fits to $n(k)$ at 300\,K for one of the seven evaluated paths.
Analogous fits along the other directions yield qualitatively similar behavior.

To connect the electron mean free path to a resistivity estimate, the 
%Boltzmann 
conductivity was evaluated in the relaxation-time approximation~\cite{robarts_extreme_2020},
\begin{equation}
  \sigma_i = \frac{e^2}{12\pi^3\hbar}\,\ell_{\mathrm{mf}, i}\,\mathcal{A}_i\,,
  \label{eq:boltzmann}
\end{equation}
where $\mathcal{A}_i$ is the Fermi-surface area of sheet $i$. All sheets contribute in parallel, $\rho = \left(\sum_i \sigma_i\right)^{-1}$.
The Fermi-surface is illustrated by the BSF at $E_F$ without thermal disorder (no AAM), 
shown in Fig.~\ref{Fig:aTi_eq_transport}(c)--(e) as plane cuts through the $\Gamma$ point. 
Four sheets are resolved: an inner, roughly ellipsoidal pocket around $\Gamma$ (Sheet~1), an outer barrel-like hole extending along $\Gamma$--A (Sheet~2), 
and two zone-corner pockets near K and H (Sheets~3 and~4), all marked in panel~(e). The overall shape agrees well 
with the Fermi surface reported by Souvatzis \textit{et~al.}~\cite{souvatzis_anomalous_2007}. 
The Fermi surface was located from the peaks of the Bloch spectral function on a mesh of 13\,000 $\mathbf{k}$ points. The resulting isosurfaces were triangulated and their areas integrated with the marching-cubes algorithm, with irreducible-wedge results scaled to the full zone using $D_{6h}$ symmetry, giving $\mathcal{A}_1 = 0.168\,(2\pi/a)^2$, $\mathcal{A}_2 = 0.840\,(2\pi/a)^2$, $\mathcal{A}_3 = 1.368\,(2\pi/a)^2$, and $\mathcal{A}_4 = 0.696\,(2\pi/a)^2$. 
The full Fermi surface is shown in the Supplemental Material.
The resulting relaxation time 
%Boltzmann 
resistivity is included in Fig.~\ref{Fig:aTi_eq_transport}(a). 
It agrees closely with the KG result, following the same temperature dependence, 
including the saturation. This confirms that the KG resistivity saturation is a direct consequence of the CPA--AAM treatment of thermal disorder, 
which limits the electronic mean free path at high lattice temperatures, consistent with the resistivity saturation found in first-principles transport studies of disordered 
transition-metal alloys~\cite{samolyuk_temperature_2018}.
\subsection{Comparison of resistivity in Ti and \TiAlloy{}}
\label{sec:results_Ti_vs_G5}

\begin{figure*}[ht!]
    \centering
    \includegraphics[width=\textwidth]{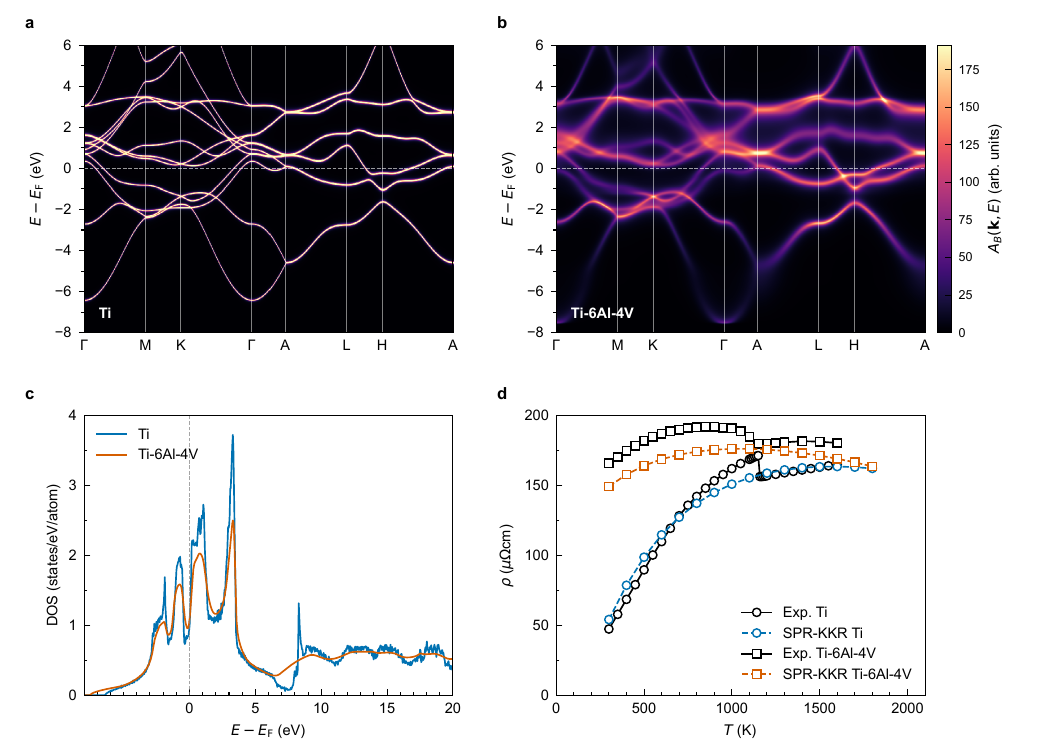}
    \caption{Comparison of electronic structure and transport between hcp Ti and hcp \TiAlloy{}.
    (a,\,b)~BSF $A_B(\mathbf{k},E)$ at 0\,K along high-symmetry paths for Ti and \TiAlloy{}, respectively. The dashed line marks $E_\mathrm{F}$.
    (c)~Total density of states per atom. The vertical dashed line indicates $E_F$.
    (d)~Temperature-dependent electrical resistivity from KKR KG calculations (colored symbols) compared with experimental data (black symbols) for Ti~\cite{belskaya_experimental_2005} and \TiAlloy{}~\cite{belskaya_emissivity_2012}.}
    \label{fig:Ti_vs_G5}
\end{figure*}

Figure~\ref{fig:Ti_vs_G5}(a,b) compares the BSF of hcp Ti and \TiAlloy{} at 0\,K. 
The band structure of Ti exhibits sharp, well-resolved bands throughout the BZ. In \TiAlloy{}, chemical disorder among 
Ti, Al, and V introduces pronounced spectral broadening across all bands, reflecting the finite quasiparticle lifetime imposed 
by the disorder within the CPA. Despite this broadening, the overall band structure is preserved. The dominant $d$-band complex near $E_\mathrm{F}$ and the 
lower-lying $sp$-derived bands remain recognizable, indicating that the alloy retains the essential electronic structure of titanium.

The primary electronic-structure effect of alloying is driven by aluminum. Although Al contributes one fewer valence electron than Ti, its $sp$ states extend the occupied manifold about $\SI{1}{\electronvolt}$ further below $E_\mathrm{F}$ and, through $sp$--$d$ hybridization, reshape the $d$-band near $E_\mathrm{F}$, as shown by Woodgate \textit{et~al.}~\cite{woodgate_emergent_2025}. Vanadium, present at only 3.6\,at.\,\%, adds further disorder scattering but has a comparatively minor effect on the band structure.

These effects also appear in the density of states, shown in Fig.~\ref{fig:Ti_vs_G5}(c). The DOS of \TiAlloy{} preserves the general shape of the Ti spectrum, but sharp features are systematically broadened, with peaks reduced in height and valleys partially filled.

Figure~\ref{fig:Ti_vs_G5}(d) shows the KG resistivity of both systems together with experimental data~\cite{belskaya_experimental_2005, belskaya_emissivity_2012}.
At low temperatures, the alloy resistivity is substantially higher than that of Ti, 
owing to a large residual resistivity of $\SI{120.3}{\micro\ohm\centi\meter}$ from chemical 
disorder~\cite{samolyuk_first_2026}, which acts as a temperature-independent elastic scattering channel. 
With increasing lattice temperature, the difference between Ti and \TiAlloy{} diminishes as thermal scattering shortens the 
electronic mean free path to a few lattice spacings, so that the additional chemical disorder contributes only marginally to the 
total resistivity. In \TiAlloy{} the hcp--bcc transition is smeared out in experiment and the rapid
resistivity drop seen in pure Ti is absent, yet
the overall trend is nonetheless reproduced by the hcp calculation.

\subsection{Electronic transport under electron--phonon nonequilibrium}

\begin{figure*}[ht!]
    \centering
    \includegraphics[width=\textwidth]{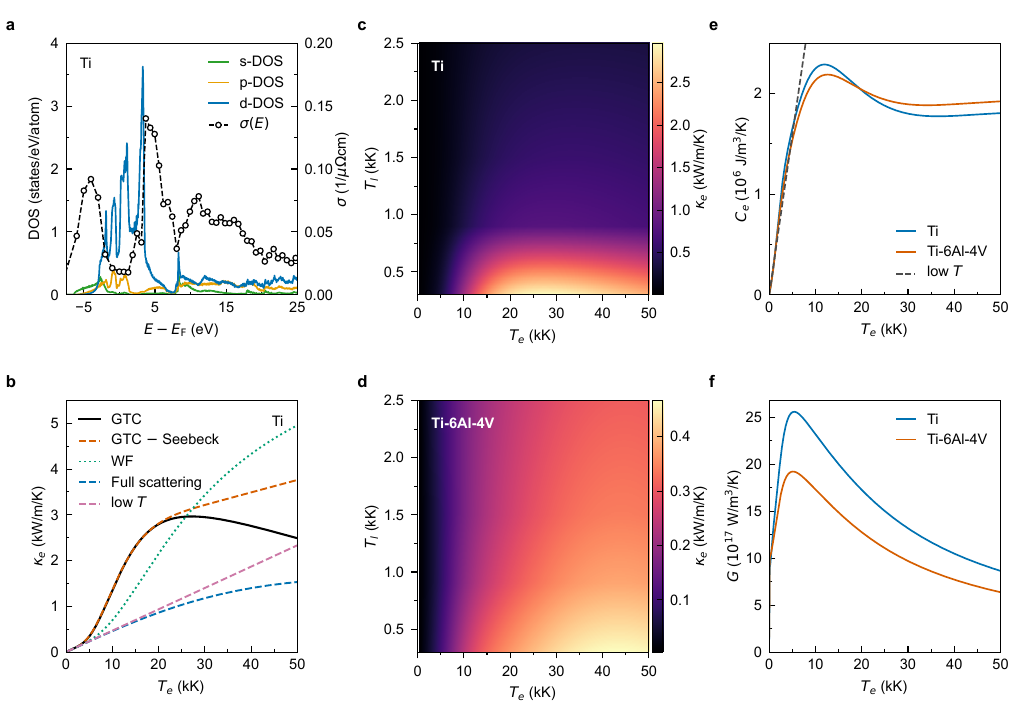}
    \caption{Electronic properties of Ti and \TiAlloy{} under electron--phonon nonequilibrium.
      (a)~Orbital-resolved density of states (solid curves, left axis) and energy-resolved conductivity $\sigma(E)$ from KG calculations at $T_{\mathrm{l}} = \SI{300}{\kelvin}$ (black circles, dashed line, right axis) for Ti; $\sigma(E)$ is suppressed where the $d$-DOS is large, as $d$ states act as scattering centers.
      (b)~Electronic thermal conductivity $\kappa_{\mathrm{e}}(T_{\mathrm{e}})$ at $T_{\mathrm{l}} = \SI{300}{\kelvin}$ for Ti: full GTC result from Eq.~(\ref{eq:kappa_e}) (black solid) against four approximations (GTC without Seebeck correction, Wiedemann--Franz, the scattering model of Eq.~(\ref{eq:rethfeld_model}), and the low-$T$ model), which agree with GTC only below about $\SI{5}{\kilo\kelvin}$, but diverge at higher electron temperatures.
      (c,\,d)~Interpolated $\kappa_{\mathrm{e}}(T_{\mathrm{e}}, T_{\mathrm{l}})$ maps for Ti and \TiAlloy{}, respectively, from GTC calculations at four lattice temperatures; the peak is a factor of $6.4$ lower in the alloy.
      (e)~Electronic heat capacity $C_{\mathrm{e}}(T_{\mathrm{e}})$ from Eq.~(\ref{eq:Ce}) for Ti and \TiAlloy{}, nearly identical for both materials.
      (f)~Electron--phonon coupling factor $G(T_{\mathrm{e}})$ from Eq.~(\ref{eq:Gep}) for Ti and \TiAlloy{}, where peak values differ by a factor of $1.3$.}
    \label{fig:kappa}
\end{figure*}

USP laser irradiation transiently heats the electronic subsystem to several thousand Kelvin, far out of equilibrium with the lattice. 
The resulting transport and coupling properties are then determined by the finite-$T_\mathrm{e}$ electronic structure.
The electronic thermal conductivity $\kappa_{\mathrm{e}}(T_{\mathrm{e}})$ is not evaluated directly but is obtained from the generalized (Onsager) transport coefficients of Eq.~(\ref{eq:GTC}), 
whose energy-resolved input is the KG conductivity $\sigma(E)$, 
computed at four lattice temperatures up to about $1.3\,T_{\mathrm{m}}$ for both materials.

Figure~\ref{fig:kappa}(a) shows $\sigma(E)$ for Ti at $T_{\mathrm{l}} = \SI{300}{\kelvin}$,
evaluated across an energy window around $E_{\mathrm{F}}$, 
extending beyond the single value $\sigma(E_{\mathrm{F}})$ that enters the equilibrium resistivity,
together with the band-resolved DOS. Near $E_{\mathrm{F}}$, where the $d$-band contribution is largest,
$\sigma(E)$ is suppressed, reflecting the role of the $d$ states as a scattering channel for the mobile $sp$ electrons rather than as a transport channel. 
Conversely, $\sigma(E)$ is large where $sp$ character dominates. It peaks at \SI{5}{\electronvolt} above $E_{\mathrm{F}}$, 
where the $d$-band weight is minimal, and remains large in the $-7.5$ to $\SI{-2.5}{\electronvolt}$ range below $E_{\mathrm{F}}$, 
before the onset of the $d$-band. Above about \SI{15}{\electronvolt} it falls off.

Figure~\ref{fig:kappa}(b) compares the resulting electronic thermal conductivity $\kappa_{\mathrm{e}}(T_{\mathrm{e}})$ at $T_{\mathrm{l}} = \SI{300}{\kelvin}$,
calculated from the full GTC expression in Eq.~(\ref{eq:kappa_e}), alongside four approximations.
The GTC result rises linearly at low $T_{\mathrm{e}}$, with a room-temperature value of 
\SI{13.5}{\watt\per\meter\per\kelvin} for Ti, passes through a broad maximum near \SI{27}{\kilo\kelvin},
and decreases at higher $T_{\mathrm{e}}$, a direct consequence of the Fermi window 
broadening into energy regions where $\sigma(E)$ decreases.
Omitting the Seebeck correction (via $\kappa_{\mathrm{e}} = L_{22}/(e^2 T_{\mathrm{e}})$~\cite{ricci_ab_2017}, orange dashed) 
overestimates $\kappa_{\mathrm{e}}$ by about 50\,\% at \SI{50}{\kilo\kelvin}, because a temperature gradient not only drives 
heat flow but also displaces charge carriers, and the resulting thermoelectric field opposes the thermal current.
The Wiedemann--Franz law $\kappa_{\mathrm{WF}} = \sigma(T_{\mathrm{e}})\, L_0\, T_{\mathrm{e}}$ (green dashed), with the 
electrical conductivity $\sigma(T_{\mathrm{e}})$ obtained from the Onsager coefficient $L_{11}$~\cite{ricci_ab_2017} 
and a constant Lorenz number $L_0$, rises monotonically to nearly twice the GTC value at \SI{50}{\kilo\kelvin}, 
making it a poor estimate at high electronic temperature.

The electronic thermal conductivity can also be modeled within the Drude picture through the electron scattering rates~\cite{rethfeld_modelling_2017},
starting from the kinetic expression $\kappa_{\mathrm{e}} = (1/3)\, c_{\mathrm{e}}\,
v^2/\nu$ with the electronic heat capacity in its low-temperature Sommerfeld form $c_{\mathrm{e}} = \gamma_{\mathrm{e}} T_{\mathrm{e}}$,
 the Fermi velocity $v = v_{\mathrm{F}}$, and the total collision frequency split into electron--electron and electron--phonon contributions,
$\nu = A_{\mathrm{ee}}\,T_{\mathrm{e}}^2 + B_{\mathrm{ep}}\,T_{\mathrm{l}}$. This yields
\begin{equation}
  \kappa_{\mathrm{e}} = \kappa_{\mathrm{e0}}\, \frac{B_{\mathrm{ep}}\, T_{\mathrm{e}}}{A_{\mathrm{ee}}\, T_{\mathrm{e}}^2 + B_{\mathrm{ep}}\, T_{\mathrm{l}}} \,.
  \label{eq:rethfeld_model}
\end{equation}
The electron--electron coefficient $A_{\mathrm{ee}}$ follows from the Quinn--Ferrell quasiparticle lifetime~\cite{quinn_electron_1958},
\begin{equation}
  A_{\mathrm{ee}} = \frac{(\pi k_{\mathrm{B}})^2}{2}\,K_{\mathrm{ee}}\,, \qquad K_{\mathrm{ee}} = \frac{\sqrt{r_\mathrm{s}}}{2\cdot 3.98\,\hbar E_{\mathrm{F}}}\,,
  \label{eq:Acoeff}
\end{equation}
where $r_\mathrm{s} = (9\pi/4)^{1/3}\,\hbar/(a_0\sqrt{2 m_{\mathrm{e}} E_{\mathrm{F}}})$ is the Wigner--Seitz radius and $a_0$ the Bohr radius. 
The $(k_{\mathrm{B}}T_{\mathrm{e}})^2$ scaling reflects the available Fermi-liquid phase space, while the $\sqrt{r_\mathrm{s}}$ dependence enters 
through the random-phase-approximation screening of the Coulomb interaction. The numerical constant $3.98$ is the Quinn--Ferrell screened-rate coefficient. 
For $E_{\mathrm{F}} = \SI{8.1}{\electronvolt}$ ($r_\mathrm{s} = 2.49$), $A_{\mathrm{ee}} = 1.36 \times 10^{6}\,$K$^{-2}\,$s$^{-1}$. The electron--phonon coefficient 
$B_{\mathrm{ep}} = \gamma_{\mathrm{e}} v_{\mathrm{F}}^2/(3\kappa_{\mathrm{e0}})$, with $v_{\mathrm{F}} = \sqrt{2 E_{\mathrm{F}}/m_{\mathrm{e}}}$ and 
$\gamma_{\mathrm{e}} = 318\,$J\,m$^{-3}$\,K$^{-2}$, is fixed by requiring Eq.~(\ref{eq:rethfeld_model}) to recover the equilibrium conductivity at 
$T_{\mathrm{e}} = T_{\mathrm{l}}$. The equilibrium value $\kappa_{\mathrm{e0}}$ is taken as the Wiedemann--Franz 
value at \SI{300}{\kelvin} (about $14\,$W\,m$^{-1}$\,K$^{-1}$). This model (blue dashed) remains well below the GTC result, reaching only about half the GTC maximum within the plotted range (its turnover lies near \SI{70}{\kilo\kelvin}, beyond the plotted range), as it assumes free-electron transport and neglects the $d$-band structure.

In the electron--phonon-dominated low-$T_{\mathrm{e}}$ limit ($A_{\mathrm{ee}}\,T_{\mathrm{e}}^2 \ll B_{\mathrm{ep}}\,T_{\mathrm{l}}$),
Eq.~(\ref{eq:rethfeld_model}) reduces to $\kappa_{\mathrm{e}} = \kappa_{\mathrm{e0}}\, T_{\mathrm{e}}/T_{\mathrm{l}}$ (pink dashed),
which rises linearly without bound. 
The electron--electron term $A_{\mathrm{ee}}\,T_{\mathrm{e}}^2$ in Eq.~(\ref{eq:rethfeld_model}) 
progressively lowers $\kappa_{\mathrm{e}}$ relative to the linear low-$T$ approximation, and a larger $A_{\mathrm{ee}}$ shifts the turnover to lower temperatures. 
Both scattering-rate models agree well with the GTC result below about \SI{5}{\kilo\kelvin}. 
Above that, a shortfall arises because they assume a constant DOS and an energy-independent free-electron conductivity, 
whereas $\sigma(E)$ rises markedly as $|E - E_{\mathrm{F}}|$ increases (Fig.~\ref{fig:kappa}(a)). 
As the Fermi window broadens with $T_{\mathrm{e}}$, it samples these higher-conductivity states and raises $\kappa_{\mathrm{e}}$ 
above the free-electron estimate, an effect captured only by the GTC integration.

Figure~\ref{fig:kappa}(c) and~(d) show the final $\kappa_{\mathrm{e}}(T_{\mathrm{e}}, T_{\mathrm{l}})$ maps used in the TTM, 
interpolated across the $(T_{\mathrm{e}}, T_{\mathrm{l}})$ plane from GTC calculations at four lattice temperatures. 
Ti reaches peak values of about $2970\,$W\,m$^{-1}$\,K$^{-1}$ at low $T_{\mathrm{l}}$ and $T_{\mathrm{e}}$ 
near \SI{27}{\kilo\kelvin}, and the peak diminishes with increasing $T_{\mathrm{l}}$ as thermal disorder reduces $\sigma(E)$. 
\TiAlloy{} peaks at only about $470\,$W\,m$^{-1}$\,K$^{-1}$ at $T_{\mathrm{e}}$ near \SI{43}{\kilo\kelvin}, a factor of 6.4 lower, 
because chemical disorder suppresses $\sigma(E)$ at all energies. The $T_{\mathrm{l}}$ dependence is also weaker in the alloy, 
since thermal scattering adds only marginally to the already dominant chemical disorder, as already seen in the equilibrium transport 
(Fig.~\ref{fig:Ti_vs_G5}(d)).

The remaining electronic structure based parameters for the TTM are the heat capacity and the electron--phonon coupling factor. 
Figure~\ref{fig:kappa}(e) shows the electronic heat capacity $C_{\mathrm{e}}(T_{\mathrm{e}})$ obtained from 
the DOS through Eq.~(\ref{eq:Ce}). For both materials $C_{\mathrm{e}}$ rises with $T_{\mathrm{e}}$ and departs 
from the low-temperature Sommerfeld form $c_{\mathrm{e}} = \gamma_{\mathrm{e}} T_{\mathrm{e}}$ as the 
Fermi window broadens into the structured $d$-band DOS, reaching a peak of 
\SI{2.29}{\mega\joule\per\cubic\meter\per\kelvin} for Ti and 
\SI{2.19}{\mega\joule\per\cubic\meter\per\kelvin} for \TiAlloy{} near \SI{12}{\kilo\kelvin}, 
before decreasing as the window extends into regions of lower spectral weight. A Sommerfeld fit gives $\gamma_{\mathrm{e}} = 318\,$J\,m$^{-3}$\,K$^{-2}$ for Ti, 
with the alloy value moderately reduced, consistent with its broadened DOS near $E_{\mathrm{F}}$. 
Figure~\ref{fig:kappa}(f) shows the electron--phonon coupling factor $G(T_{\mathrm{e}})$ evaluated through Eq.~(\ref{eq:Gep}). 
For both materials $G$ rises with $T_{\mathrm{e}}$ and reaches a peak of approximately $25 \times 10^{17}\,$W\,K$^{-1}$\,m$^{-3}$ for 
Ti and $19 \times 10^{17}\,$W\,K$^{-1}$\,m$^{-3}$ for \TiAlloy{}. The temperature dependence of both $C_{\mathrm{e}}$ and $G$ for Ti 
agrees well with the data of Lin \textit{et~al.}~\cite{lin_electron-phonon_2008}, while the absolute magnitude of $G$ is lower by a factor of about 1.5, reflecting the different $\lambda\langle\omega^2\rangle$ (a detailed comparison is given in the Supplemental Material).
While $C_\mathrm{e}$ is virtually identical for the two systems, the other electronic parameters differ significantly, 
with the peak values of $G$ and $\kappa_{\mathrm{e}}$ varying by factors of $1.3$ and $6.4$, respectively.

\subsection{Two-temperature model}
\label{sec:results_TTM}

The TTM response is evaluated at two thresholds, 
the melting threshold and the phase-explosion threshold.
The melting threshold is taken as the onset of isochoric 
homogeneous melting, $T_{\mathrm{l,max}} = 1.3\,T_{\mathrm{m}}$,
which applies in the stress-confinement regime of ultrashort 
pulses~\cite{rethfeld_ultrafast_2002}. Since the latent heat of fusion is incorporated as a Gaussian-smoothed transition in 
$C_{\mathrm{l}}(T)$ centered at $1.3\,T_{\mathrm{m}}$ (see Supplemental Material), 
the simulated temperature must exceed this nominal criterion for the heat of fusion 
to be fully absorbed, so that the effective threshold is here defined as $1.4\,T_{\mathrm{m}}$.
The phase-explosion threshold is taken as 
$T_{\mathrm{l,max}} = 0.9\,T_{\mathrm{crit}}$,
with $T_{\mathrm{crit}} = 8980\,$K the critical temperature of 
titanium~\cite{pecker_multiphase_2005}.

\begin{figure*}[!ht]
    \centering
    \includegraphics[width=\textwidth]{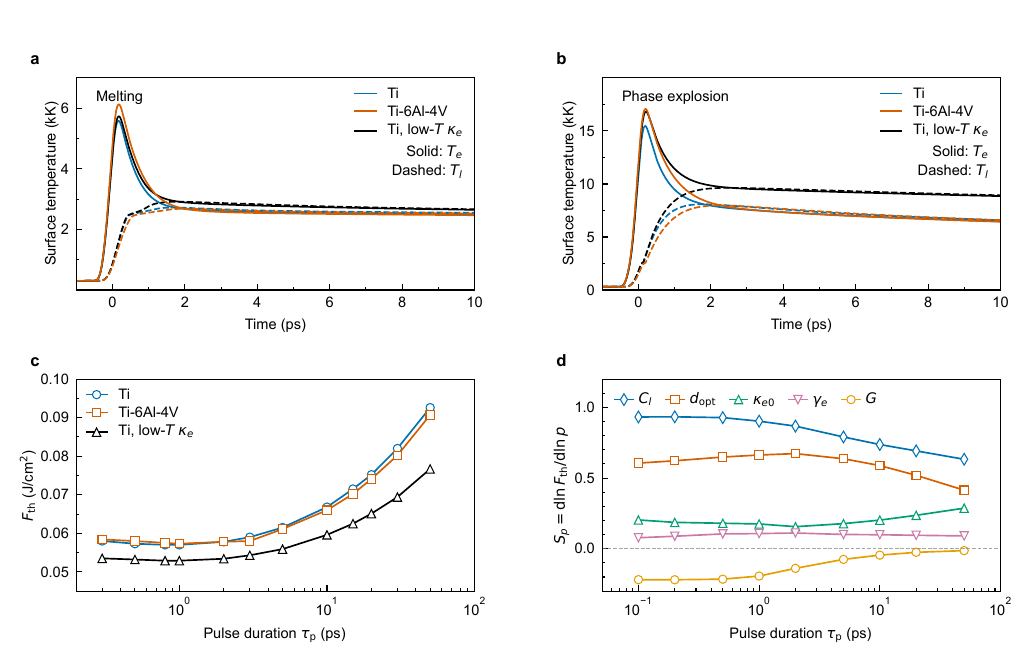}
    \caption{Two-temperature dynamics and parameter sensitivity.
  (a, b) Calculated electron and lattice temperatures, $T_{\mathrm{e}}$ and $T_{\mathrm{l}}$, for Ti,
  the simplified Ti model (with $\kappa_{\mathrm{e}} = \kappa_{\mathrm{e0}}\,T_{\mathrm{e}}/T_{\mathrm{l}}$),
  and \TiAlloy{} after ultrashort-pulse excitation, at the melting threshold (a) and the
  phase-explosion threshold (b).
  (c) Melting-threshold fluence $F_{\mathrm{th}}$ as a function of pulse duration $\tau_{\mathrm{p}}$
  for the same three parameter sets.
  (d) Logarithmic sensitivity $S_p = \mathrm{d}\ln F_{\mathrm{th}} / \mathrm{d}\ln p$ of
  the threshold fluence to the key model parameters as a function of pulse duration $\tau_{\mathrm{p}}$.
  Positive values indicate that increasing the parameter raises $F_{\mathrm{th}}$, whereas negative values indicate a reduction.}
    \label{fig:TTM}
\end{figure*}

For $\tau_\mathrm{p}=\SI{300}{\femto\second}$, the threshold fluence is determined from the Ti calculation with 
\SI{0.058}{\joule\per\centi\meter\squared} for the melting threshold 
and \SI{0.246}{\joule\per\centi\meter\squared} for the phase-explosion 
threshold, and then applied identically to \TiAlloy{} and to a simplified 
Ti model in which the electronic thermal conductivity is replaced by 
the linear low-$T$ form 
$\kappa_{\mathrm{e}} = \kappa_{\mathrm{e0}}\,T_{\mathrm{e}}/T_{\mathrm{l}}$. Figure~\ref{fig:TTM}(a) and (b) show the electron and lattice temperature 
evolution at the aforementioned thresholds. The Ti and \TiAlloy{} first-principles parameter sets yield nearly 
indistinguishable lattice heating, with $T_{\mathrm{l,max}} = 2722\,$K 
and $2683\,$K respectively, a difference of only about $1.4\,\%$. 
The alloy nonetheless reaches a higher peak electron temperature 
($T_{\mathrm{e,max}} = 6127\,$K vs.\ $5585\,$K for Ti) and equilibrates 
more slowly ($1.99\,$ps vs.\ $1.72\,$ps), reflecting its lower 
electron--phonon coupling and conductivity. The low-$T$ conductivity model yields $T_{\mathrm{l,max}} = 
2914\,$K, overshooting the first-principles result by $7.1\,\%$.
At the electron temperatures reached in this regime ($T_{\mathrm{e,max}}$ 
near $5.6\,$kK), the linear model underestimates $\kappa_{\mathrm{e}}$ 
relative to the full GTC result, reducing electronic heat diffusion into 
the bulk and concentrating energy near the irradiated surface.

A similar pattern is observed in the phase-explosion regime.
Ti and \TiAlloy{} again yield nearly identical peak lattice temperatures 
($8091\,$K and $7980\,$K), while the alloy reaches higher 
$T_{\mathrm{e,max}}$ ($17.1\,$kK vs.\ $15.5\,$kK) and equilibrates 
more slowly ($2.27\,$ps vs.\ $1.80\,$ps).
The low-$T$ conductivity model now overshoots $T_{\mathrm{l,max}}$ by $19.1\,\%$, 
reaching $9638\,$K. The larger mismatch is consistent with 
Fig.~\ref{fig:kappa}(b).
At $T_{\mathrm{e,max}}$ near $16.8\,$kK the GTC thermal conductivity 
approaches its peak while the low-$T$ approximation lies well below.

Across both regimes the conclusion is the same.
The choice of functional form for $\kappa_{\mathrm{e}}$ introduces deviations in $T_{\mathrm{l,max}}$ that are significantly larger
than the difference between the first-principles Ti and \TiAlloy{} parameter sets. The alloy-specific electronic structure modifies the electron
dynamics but leaves the lattice thermal response nearly unchanged.
This is consistent with experimental single-pulse ablation thresholds,
which for Ti ($0.28 \pm 0.02\,$J\,cm$^{-2}$ at $775\,$nm, $150\,$fs~\cite{mannion_effect_2004}) and \TiAlloy{} ($0.272 \pm 0.021\,$J\,cm$^{-2}$
at $790\,$nm, $130\,$fs~\cite{maharjan_ablation_2018}) differ by only about $3\,\%$, well within the reported measurement uncertainties.
Note that the peak lattice temperature serves only as a proxy for the ablation threshold. A quantitative comparison with measured thresholds would require an explicit ablation criterion, such as photomechanical spallation, which is beyond the scope of the present TTM.

Figure~\ref{fig:TTM}(c) shows the melting-threshold fluence $F_{\mathrm{th}}$ as a
function of pulse duration $\tau_{\mathrm{p}}$ from \SI{300}{\femto\second} to
\SI{50}{\pico\second} for the three parameter sets. Ti and \TiAlloy{} remain nearly
indistinguishable up to \SI{10}{\pico\second}. In this range $F_{\mathrm{th}}$ is
almost constant between \SI{300}{\femto\second} and \SI{3}{\pico\second}, passing
through a shallow global minimum near \SI{1}{\pico\second}, and then rises with a
square-root dependence, $F_{\mathrm{th}} \propto \sqrt{\tau_{\mathrm{p}}}$~\cite{endo_probing_2023}.
Above \SI{3}{\pico\second} the Ti threshold increases relative to the alloy, reaching
a difference of just above \SI{2}{\percent} at \SI{50}{\pico\second}.

Optimizing the pulse duration experimentally, however, requires the material response
beyond the pure TTM. Efficient ablation demands operation within the stress-confinement
regime~\cite{zhigilei_atomistic_2009,paltauf_photomechanical_2003},
$\max(\tau_{\mathrm{p}},\,\tau_{\mathrm{ep}}) < d_{\mathrm{eff}}/c_{\mathrm{l}}$,
where $c_{\mathrm{l}} = \SI{4.4}{\kilo\metre\per\second}$~\cite{casas_sound_1984} is the liquid
sound velocity and $d_{\mathrm{eff}} \approx d_{\mathrm{opt}} = \SI{20.4}{\nano\metre}$
the effective energy-deposition depth, here the optical penetration depth. This gives a
confinement time of \SI{4.6}{\pico\second}. The condition is met for both Ti and
\TiAlloy{} through the pulse duration rather than the electron--phonon equilibration
time, which is estimated from Ref.~\cite{corkum_thermal_1988} as
$\tau_{\mathrm{ep}} = \gamma_{\mathrm{e}} T_{\mathrm{e}}/G$. For the maximum electron
temperature reached here, $T_{\mathrm{e}} \approx \SI{6}{\kilo\kelvin}$ at the
\SI{300}{\femto\second} melting threshold, and with $G$ taken at its \SI{300}{\kelvin}
value, this yields $\tau_{\mathrm{ep}} \approx \SI{2}{\pico\second}$, in good agreement
with the TTM dynamics in Fig.~\ref{fig:TTM}.

Within the stress-confinement regime, ablation proceeds through spallation, which is
the most efficient removal mechanism and produces clean crater rims~\cite{redka_control_2022}.
Pulse durations below \SI{5}{\pico\second} are therefore preferable. Beyond this range
the process crosses over into the photothermal regime,
accompanied by the additional heat dissipation apparent in the TTM~\cite{ivanov_effect_2009, zhigilei_atomistic_2009, endo_probing_2023}.

Replacing the first-principles thermal conductivity with the low-$T$ approximation lowers
$F_{\mathrm{th}}$ by roughly \SI{8}{\percent} across all pulse durations below
\SI{2}{\pico\second}, consistent with the \SI{7}{\percent} higher lattice temperature at
equal fluence seen in Fig.~\ref{fig:TTM}(a), and by up to \SI{17}{\percent} at
\SI{50}{\pico\second}. The deviation is thus itself pulse-duration dependent, indicating
that the sensitivity of the threshold to the individual parameters varies with pulse
duration.

Therefore, a logarithmic sensitivity analysis of the melting-threshold fluence $F_{\mathrm{th}}$
was performed on a linearized TTM with 
$C_{\mathrm{e}} = \gamma_{\mathrm{e}} T_{\mathrm{e}}$, $\kappa_{\mathrm{e}} = \kappa_{\mathrm{e0}}\,T_{\mathrm{e}}/T_{\mathrm{l}}$,
and constant $C_{\mathrm{l}}$ and $G$.
Each parameter $p \in \{C_{\mathrm{l}},\, d_{\mathrm{opt}},\, \kappa_{\mathrm{e0}},\, \gamma_{\mathrm{e}},\, G\}$  is 
varied independently by $\pm\delta = \pm 10\,\%$ around the baseline $p_0$. The logarithmic sensitivity $S_p \equiv \mathrm{d}\ln F_{\mathrm{th}} / \mathrm{d}\ln p$ is 
approximated by the central difference $S_p = (\Delta F_+ - \Delta F_-) / (2\delta\, F_{\mathrm{th}}(p_0))$,
where $\Delta F_{\pm} = F_{\mathrm{th}}(p_0(1\pm\delta)) - F_{\mathrm{th}}(p_0)$ 
is the threshold shift when parameter $p$ is varied by $\pm\delta$ around its baseline value $p_0$.

Figure~\ref{fig:TTM}(d) shows the resulting sensitivities as a function of pulse duration. $G$ (yellow circles) and $\kappa_{\mathrm{e0}}$
(green triangles), which differ the most between the two materials in the present model, carry opposite sign.
A reduction in $G$ lowers the energy transfer rate to the lattice.
A simultaneous reduction in $\kappa_{\mathrm{e}}$ confines the deposited energy closer to the surface,
which in isolation would lower the threshold. Because both parameters are reduced in the alloy (see Fig.~\ref{fig:kappa}), the two effects largely cancel,
explaining the near-identical lattice thermal response observed above.

At short pulse durations $\tau_{\mathrm{p}}$, $G$ carries its largest electronic sensitivity, as the lattice heating rate is limited by electron--phonon energy transfer.
With increasing pulse duration the two subsystems equilibrate during the pulse 
and $|S_G|$ approaches zero.
This is consistent with $G$ not being included in the threshold condition for long pulse durations, via 
$F_{\mathrm{th}} \propto \sqrt{\kappa_{\mathrm{e0}}\, C_{\mathrm{l}}\, \tau_{\mathrm{p}}}$~\cite{chicbkov_femtosecond_1996}.
Consequently, the sensitivity of $\kappa_{e,0}$ follows the opposite trend.
It passes through a shallow minimum near $\tau_{\mathrm{p}} = 2\,$ps and rises to about $0.29$ at $50\,$ps,
as pulse-integrated thermal diffusion becomes the dominant heat-transport channel.
The cancellation between $S_G$ and $S_{\kappa_{\mathrm{e0}}}$ is therefore most effective at short pulse durations 
and weakens in the picosecond regime,
where $\kappa_{\mathrm{e0}}$ begins to dominate among the electronic parameters.
The sensitivity of $\gamma_{\mathrm{e}}$ remains nearly flat at about $0.1$ throughout, 
since the absorbed energy must be deposited into the electronic system regardless of pulse 
duration.
The lattice heat capacity $C_{\mathrm{l}}$ and the optical penetration depth $d_{\mathrm{opt}}$ dominate the absolute sensitivity at all pulse durations. 

The short-pulse limit corresponds to the analytical threshold derived by Corkum \textit{et~al.}~\cite{corkum_thermal_1988} and Wellershoff \textit{et~al.}~\cite{wellershoff_role_1999}
for an idealized delta pulse shape ($\tau_{\mathrm{p}} \rightarrow 0$) absorbed in an infinitesimal surface layer,
\begin{equation}
  F_{\mathrm{th}}^{\mathrm{Well}} = \left(\frac{128}{\pi}\right)^{1/8}
  \left(\frac{\kappa_{\mathrm{e0}}^{2}\, C_{\mathrm{l}}^{5}\, T_{\mathrm{m}}^{3}}
  {\gamma_{\mathrm{e}}\, G^{2}}\right)^{1/4},
  \label{eq:wellershoff}
\end{equation}
which yields the exponents, and thus the sensitivities of
$(S_{C_{\mathrm{l}}},\, S_{d_{\mathrm{opt}}},\, S_{\kappa_{\mathrm{e0}}},\, S_{\gamma_{\mathrm{e}}},\, S_G)_{\mathrm{Well}}$ \mbox{$= (5/4,\, 0,\, 1/2,\, -1/4,\, -1/2)$}.
The numerical sensitivities at $\tau_{\mathrm{p}} = 0.2\,$ps reproduce the ordering of the dominant exponents but are systematically reduced in magnitude,
particularly for $\kappa_{\mathrm{e0}}$ and $G$.

This deviation can be traced to the assumption of instantaneous,
pure surface absorption ($\tau_{\mathrm{p}} \to 0$, $d_{\mathrm{opt}} \to 0$) underlying Eq.~\eqref{eq:wellershoff},
which removes $d_{\mathrm{opt}}$ from the threshold by construction. In the numerical TTM, $d_{\mathrm{opt}}$ sets the dominant deposition length scale for titanium,
which transfers sensitivity away from the electronic diffusion parameters $\kappa_{\mathrm{e0}}$ and $G$ toward $d_{\mathrm{opt}}$, 
raising $S_{d_{\mathrm{opt}}}$ to about $0.62$ while reducing $|S_{\kappa_{\mathrm{e0}}}|$ and $|S_G|$ relative to Eq.~\eqref{eq:wellershoff}.

When $\tau_{\mathrm{p}} \gg \tau_{\mathrm{ep}}$,
the two subsystems equilibrate during the pulse and the TTM reduces to one-temperature heat diffusion with
$F_{\mathrm{th}} \propto \sqrt{\kappa_{\mathrm{e0}}\, C_{\mathrm{l}}\, \tau_{\mathrm{p}}}$~\cite{chicbkov_femtosecond_1996}, resulting in $S_{C_{\mathrm{l}}} \to 1/2$.
The numerical sensitivities interpolate smoothly between these two analytical limits. 
Because $|S_G|$ vanishes while $S_{\kappa_{\mathrm{e0}}}$ grows toward long pulses,
the $G$--$\kappa_{\mathrm{e0}}$ compensation that renders Ti and \TiAlloy{} nearly identical at femtosecond durations breaks down.
Here, the response becomes increasingly governed by the equilibrium thermal conductivity ($\kappa_\mathrm{e}\,(T_\mathrm{e} = T_\mathrm{l}$)).
Averaging the equilibrium thermal conductivity up to the melting range yields a ratio of $1.19$ for $\text{Ti}/\text{Ti-6Al-4V}$. 
With the long-pulse sensitivity limit of $0.5$, reached for pulse durations of several hundred \si{\pico\second} to \si{\nano\second}, Ti is then expected to
exhibit a \SI{9}{\percent} higher ablation threshold than \TiAlloy{}. The onset of this trend is already visible as the \SI{2}{\percent} difference at \SI{50}{\pico\second} from the TTM results above.
\section{Conclusion}

Alloys are central to laser micromachining and surface-functionalization applications, yet predictive ultrashort-pulse laser simulations are frequently constrained by the absence of alloy-specific electronic material parameters. This limitation is particularly evident for \TiAlloy{}, for which simulations have commonly relied on parameterizations derived from elemental Ti.

The temperature-dependent electrical resistivity, the electronic thermal conductivity, electronic heat capacity, and the electron--phonon coupling of hcp Ti and \TiAlloy{} were computed, 
providing a comprehensive alloy-specific, fully temperature-dependent electronic parameter set for 
two-temperature modeling.

For Ti, the KG resistivity reproduces the experimentally observed high-temperature saturation, 
traced to the progressive loss of quasiparticle coherence within the CPA--AAM framework. 
The alloy retains the essential band topology of Ti but disorder introduces substantial spectral broadening and a 
large residual resistivity of \SI{120.3}{\micro\ohm\centi\meter} from chemical disorder. The electronic thermal conductivity $\kappa_\mathrm{e}$, 
obtained from the full generalized (Onsager) transport coefficients, saturates and decreases at high $T_{\mathrm{e}}$ for both materials, 
peaking a factor of 6.4 lower in the alloy (about 470 versus $2970\,$W\,m$^{-1}$\,K$^{-1}$), a behavior not captured by commonly used approximations.

In TTM simulations at the melting and phase-explosion thresholds, 
the first-principles Ti and \TiAlloy{} parameter sets yield nearly identical peak lattice temperatures 
(within about $1.4\,\%$), despite a large difference in $G$, which peaks at 
$25$ versus $19 \times 10^{17}\,$W\,K$^{-1}$\,m$^{-3}$ in Ti and \TiAlloy{} respectively. 
A logarithmic sensitivity analysis dependent on pulse duration shows that the sensitivities of $G$ and $\kappa_{\mathrm{e}}$ are of comparable magnitude but opposite sign, canceling each other out. Thus, the simulated melting-threshold fluences of the two materials differ by at most about $2\,\%$ for pulse durations ranging from \SI{300}{fs} to \SI{50}{ps}, consistent with reported experimental single-pulse ablation thresholds, which agree to within about $3\,\%$.

Replacing the first-principles electronic thermal conductivity with the commonly used low temperature approximation, shifts $T_{\mathrm{l,max}}$ 
by up to $19\,\%$ for Ti. The functional form of the transport model therefore carries greater 
weight for predictive TTM accuracy in the ultrashort pulse regime than the distinction between Ti and \TiAlloy{}.

The resulting dataset provides an alloy-specific first-principles basis for calculating transient electron and lattice temperatures in \TiAlloy{} without substituting elemental-Ti transport properties. More broadly, our approach establishes a transferable framework for deriving electronic transport and thermophysical parameters of chemically disordered alloys under electron-phonon nonequilibrium at finite temperatures. Coupling such parameter sets to phase-transition, mechanical,  hydrodynamic and molecular dynamic models should enable increasingly quantitative predictions of ablation thresholds, material-removal efficiency, and processing quality in technologically relevant alloys.

%% --- AUTHOR CONTRIBUTIONS (CRediT) ---
\section*{Author Contributions}
\textbf{K.H.}: Investigation, Visualization, Writing -- original draft.
\textbf{G.E.A.}: Investigation, Writing -- review \& editing.
\textbf{A.M.}: Methodology, Writing -- review \& editing.
\textbf{M.J.V.}: Methodology, Software, Writing -- review \& editing.
\textbf{J.M.}: Methodology, Software, Writing -- review \& editing, Funding acquisition.
\textbf{H.P.H.}: Supervision, Writing -- review \& editing, Funding acquisition.
\textbf{D.R.}: Conceptualization, Investigation, Methodology, Visualization, Supervision, Writing -- original draft.

%% --- ACKNOWLEDGMENTS ---
\begin{acknowledgments}
This work was supported by Deutsche Forschungsgemeinschaft under Grant 528706678 awarded to H.P.H. and the project QM4ST (Quantum Materials for Applications in Sustainable Technology), funded as project No.\ CZ.02.01.01/00/22\_008/0004572 by the Czech Ministry of Education, Youth and Sports through Programme Johannes Amos Comenius, call Excellent Research, awarded to J.M.
This work was further supported by the Ministry of Education, Youth and Sports of the Czech Republic through the e-INFRA CZ (ID:90254). A.M. acknowledges partial financial support also from the Czech Science Foundation (GA \v{C}R) project no.~23-04746S. 
G.E.A. and M.J.V. acknowledge funding from Fonds de la Recherche Scientifique (FRS-FNRS Belgium) and Fonds Wetenschappelijk Onderzoek (FWO Belgium) for EOS project CONNECT (G.A. 40007563).
M.J.V. acknowledges funding by the Dutch Gravitation program
“Materials for the Quantum Age” (QuMat, reg number 024.005.006), financed by the Dutch Ministry of Education, Culture and Science (OCW).
The authors further gratefully acknowledge the computational and data resources as well as the support provided by the Leibniz Supercomputing Centre (\href{https://www.lrz.de}{lrz.de}). J.M. acknowledges VSB – Technical University of Ostrava, IT4Innovations National Supercomputing Center, Czech Republic, for awarding this project access to the LUMI supercomputer, owned by the EuroHPC Joint Undertaking, hosted by CSC (Finland) and the LUMI consortium through the Ministry of Education, Youth and Sports of the Czech Republic through the e-INFRA CZ (grant ID: 90254). 
G.E.A. and M.J.V. acknowledge computing time from: CECI (FRS-FNRS Belgium Grant No. 2.5020.11); the Lucia Tier-1 of the F\'ed\'eration Wallonie-Bruxelles (Walloon Region grant agreement No. 1117545); and the EINF-16505 grant ``Spin and heat transport from first principles'' on LUMI (see above) granted through SURF B.V. (NL). 
\end{acknowledgments}

%% --- DATA AVAILABILITY ---
\section*{Data Availability}
The data that support the findings of this study are available from the corresponding authors upon request.

%% --- DECLARATION ---
\section*{Declaration of Competing Interests}
The authors declare no competing interests.

\bibliography{bib}

\end{document}